\documentclass{aastex}
\usepackage{emulateapj5}

\usepackage{epsfig}
\usepackage{rotate}

\def \Teff{$T\sb{\rm eff}$}
\def \logg{$\rm log~{\it g}$}
\def \lya{Ly-$\alpha$}
\def \lyb{Ly-$\beta$}
\def \lyg{Ly-$\gamma$}
\def \fuse{{\it FUSE}}
\def \ie{i.e.}
\def \eg{e.g.}

\shorttitle{First detection of the H$_2$ quasi-molecular satellite at 
1150\,\AA\ with {\it FUSE}}
\shortauthors{Allard et al.}

\begin{document}

\title{{\it FUSE} observations of G226$-$29:\\
First detection of the H$_2$ quasi-molecular satellite at 1150\,\AA}

\author{N. F. Allard\altaffilmark{1,2}, G. H\'ebrard\altaffilmark{1}, 
J. Dupuis\altaffilmark{3}, P. Chayer\altaffilmark{3,4}, 
J. W. Kruk\altaffilmark{3}, J. Kielkopf\altaffilmark{5}, 
I.~Hubeny\altaffilmark{6}}

\altaffiltext{1}{Institut d'Astrophysique de Paris, CNRS, 
F-75014 Paris, France (allard@iap.fr, hebrard@iap.fr)}
\altaffiltext{2}{Observatoire de Paris-Meudon, LERMA, Place
Jules-Janssen, F-92195 Meudon, France}
\altaffiltext{3}{Department of Physics and Astronomy, Johns Hopkins 
University, Baltimore, MD 21218, USA (jdupuis@pha.jhu.edu, 
chayer@pha.jhu.edu, kruk@pha.jhu.edu)}
\altaffiltext{4}{Primary affiliation: Department of Physics and
Astronomy, University of Victoria, P.O. Box 3055, Victoria, BC V8W
3P6, Canada}
\altaffiltext{5}{Department of Physics, University of Louisville,
Louisville, KY 40292, USA (john@aurora.physics.louisville.edu)}
\altaffiltext{6}{NOAO, 950 North Cherry Avenue, Tucson, AZ 85726, USA 
(hubeny@noao.edu)}

\begin{abstract}

We present new FUV observations of the pulsating DA white dwarf
G226$-$29 obtained with the Far Ultraviolet Spectroscopic Explorer
(\fuse).  This ZZ~Ceti star is the brightest one of its class and the
coolest white dwarf observed by \fuse.  We report the first detection
of the broad quasi-molecular collision-induced satellite of \lyb\ at
1150~\AA, an absorption feature that is due to transitions which take
place during close collisions of hydrogen atoms.  The physical
interpretation of this feature is based on recent progress of the line
broadening theory of the far wing of \lyb. This predicted feature had 
never been observed before, even in laboratory spectra.

\end{abstract}

\keywords{line: profiles -- stars: individual (G226$-$29) --
stars: white dwarfs -- stars: atmospheres -- ultraviolet: stars}

\section{Introduction}

DA white dwarfs have hydrogen-rich atmospheres whose FUV spectra show
the Lyman series lines. Spectra of these high gravity stars are
modeled by theoretical spectra based on the Stark broadening of the
hydrogen lines (see, \eg, Vidal, Cooper, \& Smith 1973; Finley et
al. 1997; Barstow et al.~2003).  FUV and UV observations of some DA
white dwarfs show, however, strong deviations from the Stark
broadening.  Greenstein~(1980) and Holm et al.~(1985) reported that
International Ultraviolet Explorer ({\it IUE}) spectra of cool white
dwarfs show strong absorption features in the far wing of \lya\ near
1400~\AA\ and 1600~\AA.  Koester et al.  (1996) announced the presence
of similar features in the red wing of \lyb\ by analyzing observations
of Wolf~1346 obtained with the Hopkins Ultraviolet Telescope.  More
recently, absorption features in the wing of \lyg\ were detected in
the Orbiting and Retrievable Far and Extreme Ultraviolet Spectrometers
({\it ORFEUS}) and Far Ultraviolet Spectroscopic Explorer (\fuse)
spectra of white dwarfs (Koester et al.  1998; Wolff et al. 2001;
H\'ebrard et al.  2003; Dupuis et al. 2003).  All these absorption
features are interpreted as quasi-molecular satellites of \lya\ 
(Koester et al. 1985; Nelan \& Wegner 1985), \lyb\ (Koester et
al. 1996), or \lyg\ (H\'ebrard et al.~2003). Quasi-molecular lines
arise from radiative collisions of excited atomic hydrogen with
unexcited neutral hydrogen atoms or protons.

Theoretical calculations of the complete \lyb\ line profile which
include perturbations by both neutral hydrogen and protons (Allard et
al. 2000) have been used recently to improve theoretical modeling of
synthetic spectra for cool DA white dwarfs (H\'ebrard et al. 2002a).
These new calculations reveal in particular a broad H$_2$
collision-induced satellite in the red wing of \lyb\ at 1150~\AA.
This satellite appears in models covering a narrow range of effective
temperatures and is strong enough to be detected in spectra of cool DA
white dwarfs with $10000 \le T\sb{\rm eff} \le 13000$~K.  In addition
to that, the H$_2$ $\lambda$1150 satellite is very sensitive to the
degree of ionization and represents a potentially important means to
diagnose the convective mixing efficiency in DA white dwarfs,
especially in this range of effective temperatures (see, \eg, Bergeron
et al.  1992). Even in laboratory work, where a laser-produced plasma
generates controlled conditions similar to those encountered in stellar
atmospheres of cool white dwarfs, this transition has yet to be 
detected.

The effective temperature domain in which the H$_2$ $\lambda$1150
satellite is the strongest includes an important class of stars: the
ZZ~Ceti stars.  ZZ~Ceti stars are a class of hydrogen-rich atmosphere
white dwarfs exhibiting multi-periodic variations with period ranging
from 100 to 1000~s that are produced by non-radial g-modes (see, \eg,
the recent review by Fontaine, Brassard, \& Charpinet 2003).
According to Bergeron et al.  (1995) all ZZ~Ceti stars occupy an
empirical instability strip delimited by $\rm 11100 K \le T_{eff} \le
12500 K$.  These objects have a great astrophysical interest because
seismological studies allow to probe their interiors and therefore
provide insights for constraining the stellar structure and evolution
of white dwarfs.  ZZ~Ceti stars represent an evolutionary phase
through which all DA stars must evolve. H\'ebrard et al.~(2002a) have
predicted that both H$_2$ and H$_2^+$ satellites of the \lya\ {\em
and} \lyb\ lines should be detectable in ZZ~Ceti stars.

Far-UV spectra have never been explored for any ZZ~Ceti star. The
observations obtained with {\it IUE} and {\it HST} were limited to the
red wing of \lya. Here, we present a far-UV spectrum (down to
1000~\AA) of the brightest object of this class, G226$-$29.  These
data allow us to report the first detection of the H$_2$ \lyb\ 
collision-induced (CI) satellite at 1150~\AA.

\section{The collision-induced satellite at 1150~\AA}

\begin{figure}
\hspace{5cm}\psfig{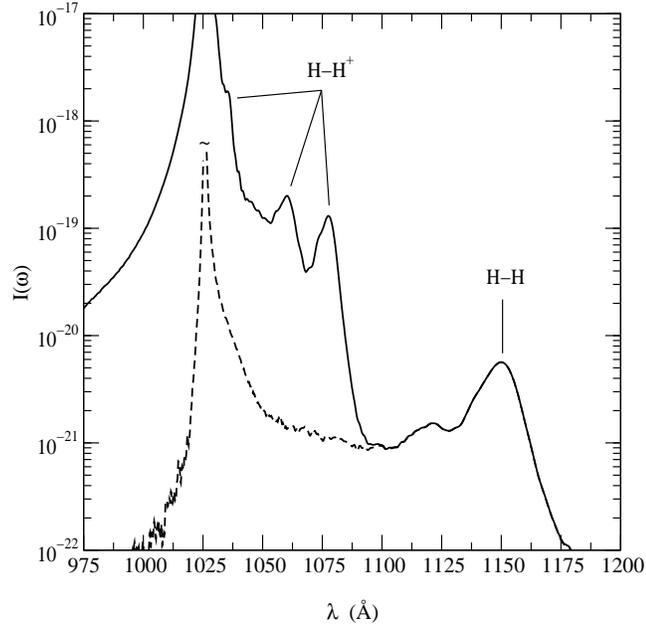}
\caption{Line profile of \lyb\ perturbed by neutral hydrogen and
protons (solid line). The dashed line shows only the contribution from
neutral perturbers. $I_\omega$ is the normalized line profile
proportional to the absorption coefficient as described by Allard et
al.~(1999).
\label{betatot}}
\end{figure}

Based on a theoretical study of the \lyb\ profile of atomic hydrogen
perturbed by collisions with neutral hydrogen atoms and protons
(Allard et al. 2000), we predict a CI satellite feature of H$_2$ at
1150~\AA. In summary, our theoretical approach is based on the quantum
theory of spectral line shapes of Baranger~(1958a, 1958b) developed in
an {\it adiabatic representation} to include the degeneracy of atomic
levels (Royer~1974; Royer~1980; Allard et al. 1994).  Key ingredients
in our calculations are the potential energies $V(R)$ for each
electronic state of the H$_2^+$ or H$_2$ molecule ($R$ denotes the
internuclear distance between the radiator and the perturber). For
H-H$^+$ collisions, we have used the H$_2^+$ potentials calculated by
Madsen \& Peek~(1971) and, for H-H, the H$_2$ potentials calculated by
Detmer et al. (1998) and Schmelcher~(2000).  The resulting complete
\lyb\ profile is shown in Fig.~\ref{betatot}.  The broad line
satellite near 1150~\AA\ is due to the perturbation in an atomic
hydrogen collision corresponding to the
$\mathrm{B}''\bar{\mathrm{B}}\;^1\Sigma^+_u -
\mathrm{X}\;^1\Sigma^+_g$ transition of H$_2$.

\begin{figure}
\hspace{5.1cm}\psfig{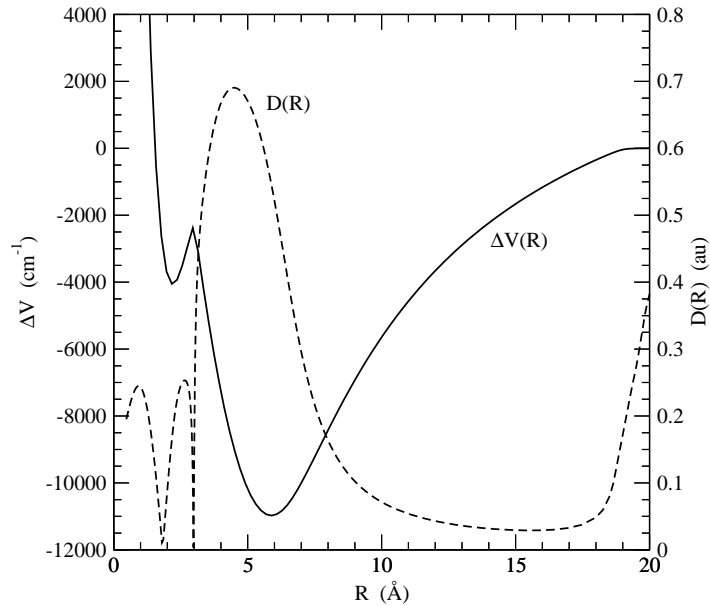}
\caption{Difference potential energy $\Delta V$ in cm$^{-1}$ and the
corresponding electric dipole moment $D(R)$ in atomic units for the
$\mathrm{B}''\bar{\mathrm{B}} - \mathrm{X}$ transition which gives the
collisional-induced satellite of \lyb.
\label{singlets}}
\end{figure}

The properties of the H$_2$ $\lambda 1150$ CI satellite can be better
understood by studying Fig.~\ref{singlets}, which illustrates the
difference potential energies $\Delta V (R)$ for this transition:
\begin{equation}
\Delta V(R)=V_{e' e}(R) = V_{e' }(R) - V_{ e}(R) \;  ,\;
\end{equation}
where $e$ and $e'$ label the energy surfaces on which the interacting
atoms approach the initial and final atomic states of the transition
as $R \rightarrow \infty$.  The most significant characteristic of the
potential curve is the existence of the double wells.  Each extremum
in the difference potential leads, in principle, to a corresponding
satellite feature in the wing of \lyb.
Another important feature is that at larger internuclear separation,
up to $R=19$~\AA, the state has an ionic character. The ionic
interaction weakens slowly ($1/R^2$ versus $1/R^6$ dependence) thus
making the potential energy difference broad compared to the steep
well of the B-X transition which gives rise to the 1600~\AA\ satellite
in the \lya\ wing. This different shape is important because the
position of the extremum and the functional dependence of the
potential difference on internuclear separation determine the
amplitude and shape of the satellites (Allard et al. 1994). 

The satellite amplitude depends also on the value of the electric
dipole transition moment $D(R)$ (dashed line on Fig.~\ref{singlets})
taken between the initial and final states of the radiative
transition. The dependence of the moment on the separation of the
atoms during a collision modifies relative contributions to the
profile along the collision trajectory.  Line profile calculations
have been done in a classical path theory which takes into account the
variation of the electric dipole moment during a collision (see,
Allard et al.~1999).  For the transitions contributing to \lya\ and
\lyb, we used the dipole moments for H$^+_2$ and H$_2$ that Ramaker \&
Peek~(1972) and Spielfiedel~(2003) calculated as a function of
internuclear distances.

The qualitative effects of the $\mathrm{B}''\bar{\mathrm{B}}\ -
\mathrm{X}$ transition can be discerned in Fig.~\ref{singlets}, 
which shows the radiative dipole transition moment and the difference
potential energies as a function of $R$.  The transition dipole moment
is small for most values of $R$ but shows a broad maximum at 4.5~\AA,
close to the broad minimum of the outer well in the potential
difference.  In contrast the dipole moment is small in the vicinity of
the inner well in the potential difference.  As a result, the main
contribution from this transition is a CI satellite in the far wing
centered at a frequency corresponding to the broad minimum (Allard and
Kielkopf~1982).

The CI absorption depends strongly on the internuclear separation and
produces a broad spectral feature with a characteristic width of the
order of the inverse of the duration of the close collision.

The line satellite shown in Fig.~\ref{betatot} presents a shoulder at
1120~\AA, a similar shape has been obtained for the 1600~\AA\ 
satellite.  In Fig.~6 of Allard~et al.~(1999) both theory and
experiment show an oscillatory structure between the satellite and the
line, with a minimum at about 1525~\AA.  These oscillations are an
interference effect (Royer~1971; Sando \& Wormhoudt~1973), and are
expected to depend on the relative velocity of the collision and
therefore on temperature.  To conclude, we emphasize the importance of
the accuracy of both the potential energies {\em and} the dipole
moments for the line shape calculations and of a theory that takes
into account the variation of the dipole moment during an atomic
collision.

\section{\fuse\ observation of G226$-$29}
\label{reduction}

G226$-$29 (WD1647+591, DN Dra) is the brightest ZZ~Ceti star. It was
originally discovered to be variable by McGraw \& Fontaine (1980) but
the first detailed analysis of the light curve had been published by
Kepler, Robinson, \& Nather (1983). This is one of the best studied
ZZ~Ceti stars, for which Fontaine et al. (1992) proposed that
pulsation properties could be interpreted assuming a ``thick''
hydrogen layer mass of $\log q({\rm H}) \approx -4.4$. Subsequent
observations by the Whole Earth Telescope consortium (Kepler et
al. 1995) and by {\it HST/FOS} (Kepler et al. 2000) allowed
unequivocal mode identification and supported Fontaine et al.'s (1992)
contention that G226$-$29 has a thick hydrogen layer.  G226$-$29
defines the blue (hot) edge of the instability strip according to
optical determination of Bergeron et al.~(1995).

G226$-$29 was observed by \fuse\ as part of our Cycle~4 Guest
Investigator Program D101. This ZZ Ceti star is the coolest white
dwarf observed with this instrument so far.  Two observations of four
exposures each were obtained on 2003 March 11 and 12 in time-tagged
photon address mode (TTAG) with the object in the 30''$\times$30''
aperture (LWRS). The total duration was 29.1~ksec ($\sim 8$~h).
Details of the \fuse\ instrument may be found in Moos et al.~(2000)
and Sahnow et al.~(2000).

The one-dimensional spectra were extracted from the two-dimensional
detector images and calibrated using version~2.4.1 of the CalFUSE
pipeline. The \fuse\ detector segments of the two observations were
co-added and projected on a $0.6$~\AA-pixel base, \ie\ pixels about
100 times larger than the original \fuse\ detectors pixels.  This
degradation of the \fuse\ spectral resolution (typically
$\lambda/\Delta\lambda\simeq15000$ for this kind of target in the
large slit; see H\'ebrard et al. 2002b; Wood et al. 2002) is of no
effect on the shapes of the large stellar features that we study and
allows us to increase the signal-to-noise ratio. At this resolution,
no spectral shifts were detected between the different co-added
spectra.

The two shorter wavelength segments (SiC1B and SiC2A,
$\sim905$\,\AA$-1000$\,\AA) were not used as no significant stellar
spectrum was detected on them. We also did not use the LiF1B segment
in our final spectrum, which is known to present a large-scale
distortion (the {\it worm}) in the flux calibration that spans the
region of interest (see, \eg, H\'ebrard et al.~2002a). Thus, the range
$1105$\,\AA$-1180$\,\AA, in which the H$_2$ satellite is detected
(Sect.~\ref{comparison}), comes from the LiF2A segment only. We
checked that the H$_2$ satellite is also detected on LiF1B, despite
the alteration of the worm.  This redundancy in the \fuse\ spectral
coverage allow us to conclude that the feature around 1150~\AA\ is
real, and not the result of an instrumental artifact.

The final reduced spectrum is plotted in Fig.~\ref{fusesat}.

\section{Synthetic spectra and comparison with the observation}
\label{comparison}

We computed a small grid of LTE model atmospheres with pure hydrogen
composition that explicitly include the \lya\ and \lyb\ 
quasi-molecular opacities.  These are determined for $T = 12000$~K,
computed from absorption profiles that take into account a variable
dipole, modulated by the Boltzmann factor (Allard et al. 1999).  The
atmosphere models have been calculated using the program TLUSTY
(Hubeny 1988; Hubeny \& Lanz 1992, 1995).  The synthetic spectra were
computed by using the spectral synthesis code SYNSPEC.  The line
satellites have a strong blanketing effect and have been included in
both the atmosphere model and synthetic spectra calculations.
Convection was treated within the usual framework of the mixing length
approximation (MLT), where we used ML2/$\alpha$=0.6 (Bergeron et
al. 1995).

\begin{figure}
\hspace{1.cm}\psfig{file=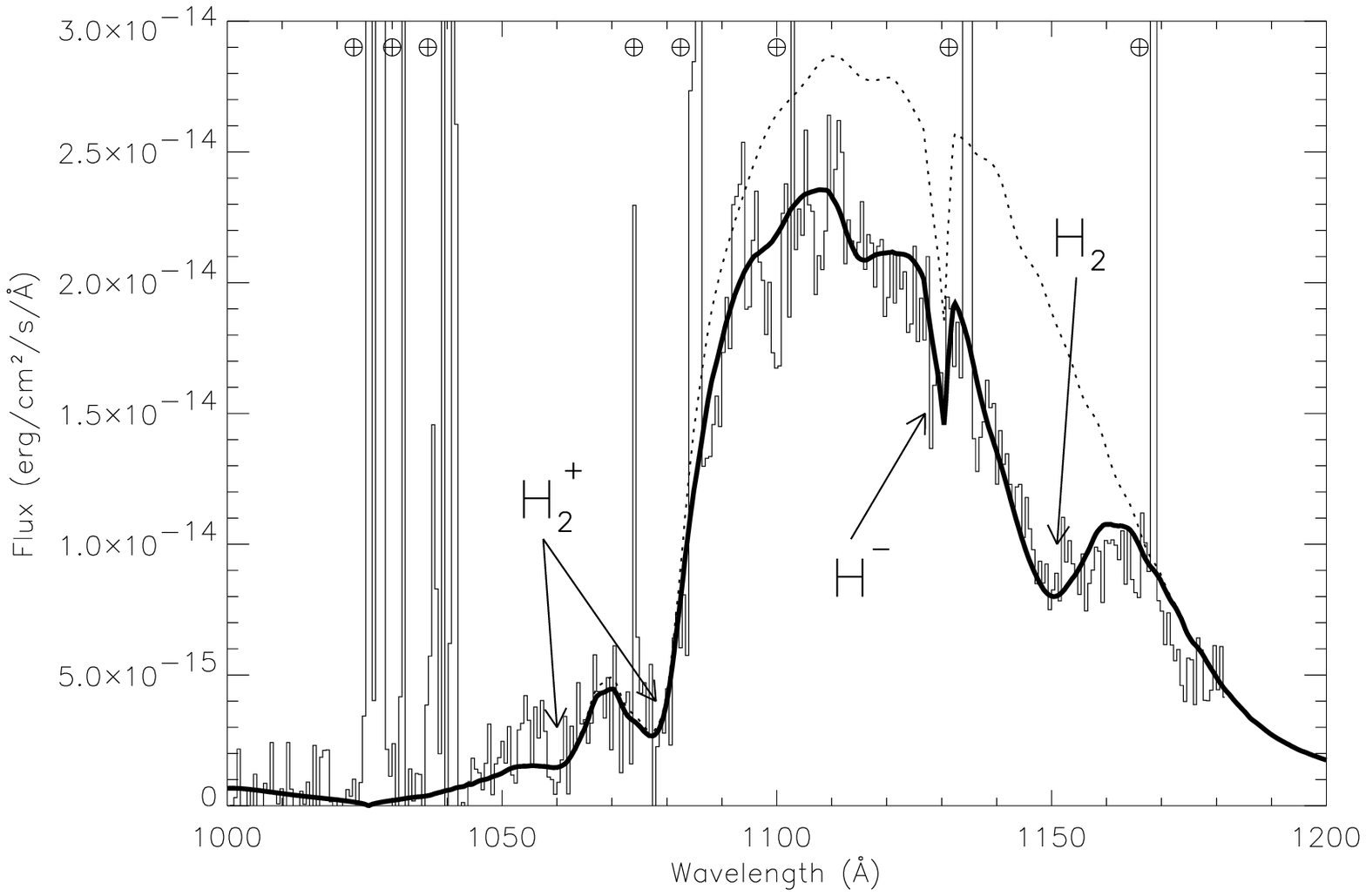,height=13cm}
\vspace{0.8cm}
\caption{\fuse\ spectrum (histogram) compared with the theoretical
model (solid line) with the new \lyb\ broadening including the
quasi-molecular satellites.  The H$_2$ quasi-molecular satellite at
1150~\AA\ is detected for the first time.  The dotted line is the
model with no H$_2$ transitions included.  The emission lines (labeled
$\earth$) are due to the airglow emission in the high terrestrial
atmosphere (Feldman et al.~2000).
\label{fusesat}}
\end{figure}

Fig.~\ref{fusesat} compares our best model to the \fuse\ spectrum of
G226$-$29. The observed feature at 1150~\AA\ is well reproduced by the
predicted H$_2$ satellite, both in terms of shape and wavelength
position. This is the first detection and identification of that
feature.  We also computed a model without including the H-H
interactions to appreciate how these interactions affect the far wing
of \lyb.  Fig.~\ref{fusesat} shows that this model (dotted line)
neither produces the H$_2$ $\lambda$1150 absorption feature nor the
significant flux absorption down to 1085~\AA.  This illustrates that
an accurate fit of the observed FUV spectra cannot be obtained without
using the improved theoretical profile calculations including
perturbations by both protons {\it and} neutral hydrogen presented by
Allard et al.~(2000).

As expected in that temperature range, the H$_2^+$ satellite at
1076~\AA\ is also detected in spite of the low flux level. The H$_2^+$
satellite at 1058~\AA\ is not clearly detected because of the too low
signal-to-noise ratio.  The H$^-$ absorption at 1130~\AA\ 
(Wishart~1979) is detected; that H$^-$ feature was already detected in
the \fuse\ spectrum of the cool DA white dwarf G\,231$-$40 (H\'ebrard
et al.~2002a).

The best fit is obtained for \Teff\,$=12000$~K and \logg\,$=7.9$.
That temperature is slightly lower than the one that Bergeron et al.
(1995) determined from optical observations (\Teff\,$=12460\pm230$~K),
and it is similar to the one reported by Koester \& Holberg~(2001),
who fitted the {\it HST/FOS} data by adding the constraints of the V
magnitude and the trigonometric parallax (\Teff\,$= 12020$~K).  Our
gravity, however, is significantly lower than those obtained in both
studies (\logg\,$=8.2-8.3$). The causes of this difference are not
fully understood yet. We are now investigating these causes and hope
to resolve the discrepancy between our value of the gravity and those
of Bergeron et al.  (1995) and Koester \& Holberg (2001).  This work
is beyond the scope of the present paper and will be presented
elsewhere.  In any case, that does not jeopardize our detection and
identification of the quasi-molecular lines reported in this paper.

The model plotted in Fig.~\ref{fusesat} was normalized by a free
parameter in order to fit the \fuse\ spectrum. With that
normalization, our model fits as well the shape of the {\it IUE} data
of G226$-$29, at longer wavelengths ($\sim1250$\,\AA$-2000$\,\AA).
However, we should point out that we need to divide the {\it IUE} flux
by a factor of $\sim1.3$ to bring it to the model. Small uncertainties
in the background subtraction or misalignment of the target on the
slit during the \fuse\ and {\it IUE} observations or both, could
account for the flux discrepancy.  It is worth mentioning that
H\'ebrard et al.~(2002a) fitted the \fuse\ and {\it IUE} data of the
cool DA white dwarf G231$-$40 without applying any correction factors
to the fluxes.  Given that the flux of G231$-$40 is more than 20 times
higher than the flux of G226$-$29, systematic uncertainties in the
background subtraction are significantly reduced.  It is also easier
to quantify the alignment of the target on the slit as a function of
time and maintain a reasonable photometric accuracy for a relatively
bright target such as G231$-$40.  We considered that a difference by a
factor $\sim1.3$ in the flux calibration of observations performed
with 2 different instruments of a faint target is acceptable.  Our
model with \Teff\,$=12000$~K and \logg\,$=7.9$ fits as well the UV
spectra ($\sim1250$\,\AA$-2500$\,\AA) obtained with {\it HST/FOS} and
presented by Kepler et al.~(2000). We note that this {\it HST/FOS} 
spectrum was normalized in order to fit the {\it IUE} spectra.

\section{Conclusions}

We have presented the first far-UV spectrum of a ZZ Ceti star, namely
G226$-$29, performed with \fuse. We fitted these data using
theoretical calculations of stellar profiles that include
perturbations by both neutral hydrogen and protons. The observed
spectrum is well fitted, allowing us to reproduce the global shape of
the stellar continuum, together with the H$^-$ absorption at 1130~\AA,
the H$_2^+$ quasi-molecular satellite 1076~\AA\ (and marginally the
one at 1058~\AA), and a broad H$_2$ CI satellite at 1150~\AA. This is
the first detection and identification of that H$_2$ quasi-molecular
satellite of \lyb. This feature has been predicted but never observed
up to now.  The quality of the fit allows that absorption feature to
be unambiguously identified.

This discovery confirms our theoretical prediction and gives
confidence concerning the accuracy of the fundamental molecular data
we used.  In combination with existing optical and ultraviolet data,
fits using such models will be used in the future to allow stellar
parameters (temperature, gravity) to be constrained more accurately,
and will provide insights crucial in the modeling of cool white dwarf
atmospheres.

\acknowledgments
This work is based on observations made with the NASA-CNES-CSA Far
Ultraviolet Spectroscopic Explorer. FUSE is operated for NASA by the
Johns Hopkins University.  Financial support to U.S. participants has
been provided by NASA contracts NAS5--32985, NAG5--13714, NAG5--13715,
and NAG5--11844 (JD).  French participants are supported by CNES.

\clearpage

\begin{references}
\reference{}
Allard, N. F., \& Kielkopf, J. F. 1982, Rev. Modern Phys., 54, 1103
\reference{}
Allard, N. F., Koester, D., Feautrier, N., \& Spielfiedel, A. 1994, 
\aaps, 108, 417
\reference{}
Allard, N. F., Royer, A., Kielkopf, J. F., \& Feautrier, N.
1999, Phys. Rev. A, 60, 1021
\reference{}
Allard,~N. F., Kielkopf,~J. F., Drira,~I., \& Schmelcher,~P. 2000,
Eur. Phys. J. D, 12, 263
\reference{}
Baranger, M. 1958a,  Phys. Rev., 111, 481 
\reference{}
Baranger, M. 1958b, Phys. Rev., 111, 494
\reference{}
Barstow, M. A., Good, S. A., Burleigh, M. R., Hubeny, I., Holberg, J. B., 
\& Levan, A. J. 2003, \mnras, 344, 562
\reference{}
Bergeron, P., Wesemael, F., \& Fontaine, G. 1992, \apj, 387, 288
\reference{}
Bergeron, P., Wesemael, F., Lamontagne, R., Fontaine, G., Saffer, R.,
\&  Allard, N. F. 1995, \apj, 449, 258
\reference{}
Detmer, T., Schmelcher, P., Cederbaum, L. S. 1998, J. Chem. Phys., 
109, 9694 
\reference{}
Dupuis, J., Chayer, P., Vennes, S., Allard, N. F., \& H\'ebrard, G.
2003, \apj, 598, 486
\reference{}
Feldman, P. D., Sahnow, D. J., Kruk, J. W., Murphy, E. M., \& Moos, 
H. W. 2001, J. Geophys. Res., 106, 8119
\reference{}
Finley, D. S., Koester, D., Kruk, J., Kimble, R. A., \& Allard, N. F. 
1997, in White Dwarfs: Proc. 10th European Workshop on White Dwarfs, 
ed. J. Isern, M. Hernanz, \& E. Garcia-Berro (Dordrecht: Kluwer), 245
\reference{}
Fontaine, G., et al. 1992, \apj, 399, L91
\reference{}
Fontaine, G., Brassard, P., \& Charpinet, S. 2003, \apss, 284, 257
\reference{}
Greenstein, J. L. 1980, \apj, 241, L89
\reference{}
H\'ebrard, G., Allard, N., Hubeny,~I, Lacour,~S., Ferlet,~R., \&
Vidal-Madjar,~A. 2002a, \aap, 394,~647
\reference{}
H\'ebrard, G., et al. 2002b, \apjs, 140, 103
\reference{}
H\'ebrard, G., Allard, N. F., Kielkopf, J. F., Chayer, P., Dupuis, J.,
Kruk, J. W., \& Hubeny, I. 2003, \aap, 405, 1153
\reference{}
Holm, A.~V., et al. 1985, \apj, 289, 774 
\reference{}
Hubeny, I. 1988, Comp. Phys. Comm., 52, 103
\reference{}
Hubeny, I., \& Lanz, T. 1992, \aap, 262, 501
\reference{}
Hubeny, I., \& Lanz, T. 1995, \apj, 439, 875
\reference{}
Kepler, S. O., Robinson, E. L., \& Nather, R. E. 1983, \apj, 271, 544
\reference{}
Kepler, S. O., et al. 1995, \apj, 447, 874
\reference{}
Kepler, S. O., et al. 2000, \apj, 539, 379
\reference{}
Koester, D., Weidemann, V., Zeidler-K.T., E.-M., \& Vauclair, G. 
1985, \aap, 142, L5
\reference{} 
Koester, D., Finley, D.~S., Allard, N. F., Kruk, J. W., \& Kimble,
R. A. 1996, \apj, 463, L93
\reference{}
Koester,~D., Sperhake,~U., Allard,~N. F., Finley,~D. S., \& Jordan,~S. 
1998, \aap, 336, 276 
\reference{}
Koester, D., \& Holberg J. B. 2001, in ASP Conf. Ser. 226, 12th European
Workshop on White Dwarfs, ed. J. L. Provencal, H. L. Shipman,
J. MacDonald, \& S. Goodchild (San Francisco:ASP), 299
\reference{} 
Madsen, M. M., Peek, J. M. 1971,  Atomic Data, 2, 171
\reference{} 
McGraw, J. T., \& Fontaine, G. 1980, unpublished
\reference{} 
Moos, H. W., et al. 2000, \apj, 538, L1
\reference{} 
Nelan, E. P., \& Wegner, G. 1985, \apj, 289, L31
\reference{}
Royer, A. 1971, Phys. Rev. A, 43, 499
\reference{}
Royer, A. 1974, Can. J. Phys., 52, 1816
\reference{}
Royer, A. 1980, Phys. Rev. A, 22, 1625
\reference{} 
Ramaker, D. E., \& Peek, J. M. 1972, J. Phys. B, 5, 2175
\reference{} 
Sahnow, D.  J., et al. 2000, \apj, 538, L7 
\reference{}
Sando,~K.~M., \& Wormhoudt,~J.~G. 1973, Phys. Rev. A, 7, 1889 
\reference{}
Schmelcher, P. 2000, Private communication
\reference{}
Spielfiedel, A. 2003,   J. Mol. Spectrosc., 217, 162
\reference{}
Vidal, C. R., Cooper, J.,\& Smith, E. W. 1973, \apjs, 25, 37
\reference{}
Wishart, A. W. 1979, MNRAS 187, 59
\reference{}
Wolff,~B., et al. 2001, \aap, 373, 674
\reference{}
Wood, B. E., et al. 2002, \apjs, 140, 91
\end{references}
\end{document}